\documentclass[english]{article}
\usepackage{jcappub}
\usepackage{graphicx}
\usepackage{appendix}
\usepackage{dcolumn}
\usepackage{caption}
\usepackage{subfigure}
\usepackage{float}
\usepackage{bm}
\usepackage{latexsym}
\usepackage{epsfig}

\begin{document}
\title{Mach's Principle and a new theory of gravitation}
\author{Santanu Das}
\emailAdd{santanud@iucaa.ernet.in}
\affiliation{IUCAA, Post Bag 4, Ganeshkhind, Pune 411 007, India}
\date{\today}

\abstract{
Einstein was highly fascinated by Ernest Mach's work and by formulating the general theory of relativity
(GR) he tried to provide a mathematical description to the Mach's principle. 
However, soon after its formulation, it was realized that
GR does not follow Mach's principle. As it accurately explained many observational results,
Einstein did not make any farther attempt to reformulate the theory to
explain Mach's principle. Later on, several attempts were made
by different researchers to formulate a theory of gravity based on Mach's principle.
However, all these theories have their own merits and drawbacks.
In this paper a new theory of gravity is proposed that is
completely based on the Mach's principle. 
It is a metric theory and can be derived from action principle,
which guarantees to follow all the conservation laws. 

We present few examples to show that the theory  can explain the galactic rotation curves %velocity profiles
very accurately. It also explains the discrepancy between dynamic mass and the photometric mass
of the galaxy clusters without any exotic dark matter. The bullet cluster weak lensing phenomenon 
that is considered to be a strong evidence for dark matter can also be explained using this new gravitational theory.
Hence, the theory provides a viable alternative to standard gravity theory.
}

\maketitle

\section{Introduction}

Newtonian gravity, can provide a very accurate description of gravity, provided
the gravitational field is week, not time varying and the 
concerned velocities are much less than $c$. It can very accurately 
describes the motions of planets and the satellites in the solar system. 
The general theory of relativity is designed to follow Newtonian
gravity at large scale. Thought the Newtonian gravity and general relativity
precisely explain the nature of gravity at solar system scale, 
they fail to produce the galactic velocity profiles, 
provided calculations are made just considering the visible matters in the galaxy. 
This lead researchers to postulate a new form
of weakly interacting matter, named as dark matter. 
SUSY particles are considered to be the potential candidate for dark matter. 
However, present studies show that the possibility of existence of SUSY particles is 
negligible. Therefore, it is important to challenge the theory 
instead postulating the ad hoc form of matter based on a blind faith on 
Newtonian Mechanics. 

Several theories are proposed in last decade to explain the galactic
rotation curves. 
Empirical theories like MOND, \cite{Milgrim1983,Milgrim1983a,Milgrim1983b,Milgrom2011} can
explain the galactic rotation curves well, but violates
momentum conservation principles. Bekenstein proposed AQUAL \cite{Bekenstein1984,Bekenstein2009,Milgrom1986}
to provide a physical
ground to MOND. Other theories like Modified gravity (VeTeS)
\cite{Moffat2005,Brownstein2005,Brownstein2005a,Moffat2005a}, 
TeVeS \cite{Bekenstein2005}, Massive gravity \cite{Dam1970,Zakharov1970,Babichev2010,Babichev2013} 
etc. are also proposed 
to match the galactic velocity profiles without dark matter. However most of these theories 
fails to explain the Mach's principle. 

There are also theories postulated in the last centuries 
that directly follows from Mach's principle. 
Amongst those Brans Dicke theory \cite{Brans1961} and Hoyle Narlikar theory \cite{Hoyle1964}
are most notable. However, these theories requires extra dark 
matter candidates to explain the galactic velocity profiles. There are 
several other theories such as induced matter theory \cite{Overduin1998,Ponce1993,Wesson1992}, 
tired light hypothesis \cite{Accardi1995} etc. proposed by researchers.
However, none of these theories can provide an explanation for galactic velocity
profile and Mach Principle simultaneously. Therefore, a new theory of gravity
is proposed here to fill the gap. In this paper we refer this theory as Machian Gravity. 
The theory is based on the following premises

\begin{itemize}
\item Action principle : The theory should be derived
from an action principle, because that guarantees
all the conservation laws, such as the energy, momentum conservations.
\item Equivalence principle : 
The inertial properties of a particle is not completely 
its intrinsic property and depends on the nature of background.
Weak Equivalence Principle does not consider the background contribution on mass of a particle \cite{Das2012}. 
Therefore, the proposed theory is not designed to follow the WEP.

\item Departure from Newtonian gravity : As Newtonian gravity and GR are tested
several times at the solar system scale, the proposed
theory should follow them in solar system scale.
However, it deviates from the standard theories at galactic scale, 
where the standard theories fails to provide correct predictions.

\item Mach's principle : The theory is designed to follow the Mach's
principle considering that Mach's principle provides the appropriate
description of inertia. It is considered that the
inertial mass of an object is not an intrinsic property of the object
and gets its contribution from all other particles in the universe. %According to ,  
Mach's Principle can be described by taking
a five dimensional coordinate system \cite{Das2012}. Hence, in this paper same 5 dimensional
coordinate system is used to formulate the theory of gravity.

\end{itemize}
~

The paper is organized as follows. The second section provides the logical description
of the theory. In the third section, some of the mathematical tools, which 
are used for formulating the theory, are discussed. Here we also present  the source free field equations for the theory. 
The fourth section describes the field equations calculated in presence of source terms.
The static spherically symmetric vacuum solution for the theory in weak field approximation is presented in the 
fifth section. We show that the solution follows the Newtonian gravity at smaller scale but deviates 
from Newtonian gravity at large scale.
The sixth section provides some examples of galactic rotation curves and galaxy cluster mass distribution to show that the theory
can provide results that matches with the observations very accurately. The final section is the conclusion and discussion section.

\section{Logical description of the theory}

According to Mach's principle, inertial properties of matter
depend on the background. Therefore, if two identical objects are
kept at two different locations in the universe then depending on their backgrounds
the inertial masses of those two particles may be different.
In \cite{Das2012} a new formalism is proposed to explain Mach's principle by adding 
an extra dimension. The kinetic energy related to
a new coordinate dimension, labeled as $\zeta$, keeps on changing depending on the 
position of the particle in the universe and the background of the particle. 

In this section we logically describe how the Mach's principle can be related with 
of coordinate curvature. We also discuss how the $5$ 
dimensional coordinate system can provide the perfect Riemannian curvature to describe 
the Mach's hypothesis.

\subsection{Mach's principle}
\label{sec2}
The velocity or acceleration of a particle are relative measurements, i.e. 
they are always measured with respect to some reference frame \cite{Hoyle1980}. 
While measuring the velocity or the acceleration of a running train, they 
are measured with respect to the surface of the earth. But, the earth
is orbiting the sun, which is again circling our galaxy. The galaxy
also has some random motion inside the galaxy cluster and so on.
Therefore, if the origin of the coordinate system for measuring 
the velocity and the acceleration of the train is chosen to be at
the center of the galaxy then the velocity and the acceleration of the train will
be completely different. Hence choice of a perfect coordinate system is extremely necessary.

Let us consider that a stone is tied with a string and whirled around in a circle. We define
two reference frames, one with origin at the center of the circle, which
is fixed with respect to us and the other is fixed at the 
stone. We can analyze the forces on the stone, using
the Newton's law, in the reference frame that is fixed at the center of the circle.
If $m_{i}$ is the inertial mass, $v$ is the velocity
of the stone, $r$ is the radius of the circle and $T$ is the tension
on the string then using Newton's law, we can write 

\begin{equation}
m_{i}\frac{v^{2}}{r}=T\,.\label{eq:stone}
\end{equation}

Same analysis can be done in the second reference frame 
that has its origin fixed at the stone. In this reference frame, as the stone
is at rest, $v=0$. Therefore, the left hand side of
Eq.(\ref{eq:stone}) becomes zero. However, the right hand side
which is actually representing the force on the string towards the
center, remains as $T$. Therefore, the equality of the
Eq.(\ref{eq:stone}) does not hold in this frame. Newton's laws are valid
only in reference frames that have no acceleration i.e. the inertial frame. 
Newton was well aware of this fact and postulated some fictitious forces that arise
in any non inertial frame to balance the equations. These forces are
usually called the inertial forces, and these have no existence outside 
mathematics. In this example, the fictitious force 
is the centrifugal force. Its magnitude is equal and opposite
to the force $T$, i.e. $-T$, and it helps us to balance
the equations. 

However, as the acceleration is a relative quantity there is no way to define an
inertial reference frame 
because there is no other frame based on which we can calculate its acceleration. 
Surface of Earth is not an inertial frame in the true sense. Newton's law is not thus 
fully applicable in that frame and the expression used in Eq.(\ref{eq:stone}) is not
accurate as different inertial forces due to the motion of Earth and Sun etc. act on this reference
frame. Quantitatively these inertial forces are so negligible that even 
without considering these forces one can come up to a fairly
good approximation for the motion of the stone. 

However, question about fixing an inertial coordinate system remains. To overcome this problem,
Mach \cite{Jammer} proposed that the inertial reference frame can be fixed by
measuring the acceleration of the frame with respect to the distant stars (or other objects). 
Hence, distant objects somehow affect the inertial properties of matter.

\subsection{General relativity a brief description}

Einstein tried to propose General theory of Relativity (GR) in view of Mach's principle. 
Logic behind Einstein's GR is as follows. According
to the Newton's gravity, it is known that if we place a particle
in a gravitational field then 

The inertial mass of the particle $\times$ acceleration of the mass
= passive gravitational mass $\times$ strength of the gravitational
field at the place

Considering the inertial mass to be equal to the passive gravitational
mass the conclusion was that in a small region of space-time it is not
possible to distinguish between the acceleration and the gravitational
field, which is Einstein's Equivalence
principle. Next for deriving the general theory of relativity, 
showed the acceleration can be related with the curvature of space-time. 
This is shown using the following thought experiment.  

Let consider two  
systems $K$ and $K'$, whose $z$ axises are aligned.  
$K$ is inertial frame and $K'$ is a non-inertial frame that is 
rotating with constant angular velocity with respect to $K$ frame.
We consider a circle on the $x'y'$ plane of the $K'$ reference frame 
and fill its diameter and perimeter with small rigid rods of length $l$.
He shows that in $K$ frame perimeter to diameter ratio is $\pi$ as usual 
but in $K'$ frame is less than $\pi$ due to the length construction from the 
special relativity. 

He argued that this ratio can't be explained by Euclidean geometry 
and needs the concept of space-time curvature from 
the Riemannian geometry. So acceleration can be explained using 
space-time curvature. As the acceleration and gravitational
field cannot be distinguished within a small enough region of space-time,
in GR the space-time curvature is related with 
the stress energy tensor to formulate the theory of gravitation.

\subsection{Rotation of background and Mach's principle}

In the previous thought experiment Einstein considered that 
$K$ is an inertial reference frame. However, there is no way to define the inertial 
reference frame unless we consider Mach's proposal of measuring
the acceleration with respect to distant stars. This leads to a conceptual problem
in the thought experiment. If there is an observer in the $K$ frame she will 
see that the entire background created by the distant stars and galaxies etc. is static.
However, the observer in the $K'$ reference frame, finds that the 
background stars and galaxies etc are rotating in her reference frame. Therefore, 
according to Mach's principle, just by observing the rotational motion of the
background objects she can conclude that Newtons laws or Euclidean geometry are not 
applicable to her coordinate frame because she has some acceleration. 

Let us extend the above thought experiment. 
Suppose, the entire background starts rotating in such a way that the
observer in the $K'$ reference frame finds herself in a static position
with respect to the distant stars and the observer 
in the $K$ reference frame 
finds that background stars starts rotating in her reference frame. 
Therefore, even though the observers at the $K$ or
$K'$ reference frame have done nothing, her definition
for the inertial reference frame gets flipped. According to Machian
concept about an inertial frame, in this present situation the $K'$
reference frame will be an inertial frame and $K$ will become a non
inertial reference frame. Hence, the Newton's laws and Euclidean
Geometry will hold in the $K'$ frame but not in $K$.

Therefore, some mechanism is required to connect the background motion to the curvature 
of the coordinate system. As Einstein's
GR does not posses any such term, it cannot explain the phenomenon discussed above. 
Background contribution can be incorporated by adding a fifth dimension
($\zeta$) that is a quantifier of the background. 

\section{Mach's principle and the field equations in absence of source terms}

A line element in $5$D coordinate system can be written as $ds^{2}=\tilde{g}_{AB}dx^{A}dx^{B}$ 
where, $\tilde{g}_{AB}$ is the 5 dimensional metric. We use the indices $A, B, C, \ldots$
for representing the $5D$ coordinate system i.e. they run from $0$ to $4$ and indices 
$\alpha, \beta, \gamma, \ldots$ for representing the $4D$ system i.e. these indices run
from $0$ to $3$. We also use tilde to denote the quantities in the $5$ dimensional frame.

The 5D Minkowski the line element is taken as \cite{Das2012} 

\begin{equation}
ds^{2}=c^{2}dt^{2}-dx^{2}-dy^{2}-dz^{2}-\frac{\hbar^{2}}{4}d\zeta^{2}\,.
\end{equation}

\noindent Here all the elements of $\tilde{g}_{AB}$ being constants, 
the five dimensional Riemann Tensor,
$\tilde{R}_{BCD}^{A}$ vanishes. We consider that in absence of any gravitating body, 
the $5$ dimensional space-time remains flat. In that case we can get similar field 
equation as that of GR but in $5$D. 
Though it may look similar to Einstein's equation,
but it has several physical implications.
Vanishing of $\tilde{R}_{BCD}^{A}$ does
not imply a vanishing $R_{\nu\alpha\beta}^{\mu}$. In a non-inertial
reference frame $R_{\nu\alpha\beta}^{\mu}$ will not vanish even though
$\tilde{R}_{BCD}^{A}$ vanishes, and the non-vanishing $R_{\nu\alpha\beta}^{\mu}$
take care of all the inertial forces. In case of the example given in Sec.$\ref{sec2}$
the $K$ frame is non-inertial. So the background motion will give the non-inertial 
terms which can give the centrifugal force acting on the object and hence 
no need to add any pseudo force that has only mathematical existence 
but no physical origin. 

It is also clear from the equations that if there is any gravitating
object then it can curve the five dimensional coordinate system. Hence,
that will give a nonzero $\tilde{R}_{BCD}^{A}$, and that 
cannot be taken away using any choice of coordinate system.
But even in presence of gravitating field it may be possible to make
the $R^\mu_{\nu\alpha\beta} =0$, i.e. a 4D flat space time.

The vanishing Riemann curvature
tensor implies that the Einstein tensor also vanish, i.e.

\begin{equation}
\tilde{G}_{AB}=0\,.
\end{equation}

\noindent This is taken as the field equations for the theory  in absence of any source terms. The motivation
behind taking this particular condition as the field equation 
is same as GR, i.e. it has some special properties like it 
is the 2nd rank tensor and consists of second derivative of space-time etc.  

\subsection{Deformation of space-time from the background}

The logic, just described can be written in a more mathematical way as follows. 
The five dimensional metric can be written in the 4+1 dimensional
form as 

\begin{equation}
\tilde{g}_{AB}=\left(\begin{array}{cc}
g_{\alpha\beta}+\kappa^{2}\phi^{2}A_{\alpha}A_{\beta} & \kappa\phi^{2}A_{\alpha}\\
\kappa\phi^{2}A_{\beta} & \phi^{2}
\end{array}\right)\,,\label{eq:metric}
\end{equation}

\noindent were $g_{\alpha\beta}$ is a 4 dimensional metric, $A_{\alpha}$ is
a 4 dimensional vector and $\phi$ is a scalar.

Few straight forward calculations can show that 
$\tilde{G}_{AB}=0$, in 5 dimension translate to the following equations in
the 4 dimension

\begin{equation}
G_{\alpha\beta}=\frac{\kappa^{2}\phi^{2}}{2}\left(g_{\alpha\beta}F_{\gamma\delta}F^{\gamma\delta}/4-F_{\alpha}^{\gamma}F_{\beta\gamma}\right)-\frac{1}{\phi}\left[\nabla_{\alpha}\left(\partial_{\beta}\phi\right)-
g_{\alpha\beta}\square\phi\right]+P_{\alpha\beta}\,,
\label{eq:Einstein's tensor}
\end{equation}

\begin{equation}
\nabla^{\alpha}F_{\alpha\beta}=-3\frac{\partial^{\alpha}\phi}{\phi}F_{\alpha\beta}+Q_{\beta}\,,
\end{equation}

\begin{equation}
\square\phi=\frac{k^{2}\phi^{3}}{4}F_{\alpha\beta}F^{\alpha\beta}+U\,.
\end{equation}

\noindent Here $F_{\alpha\beta}=A_{\alpha;\beta}-A_{\beta;\alpha}$ is the field
tensor. $P_{\alpha\beta}$, $Q_{\beta}$ and $U$ are the terms containing
the derivatives of the metric elements with respect to the fifth dimension
i.e. $\zeta$.

In Eq.(\ref{eq:Einstein's tensor}), $G_{\alpha\beta}$ is a four dimensional 
Einstein's tensor used 
in 4 dimensional gravity. But as the
calculations are done in 5 dimension, the 4 dimensional
Einstein's tensor comes up with some terms in the right hand side,
even in absence of any matter in the space-time.  
These extra terms can behave in exactly the same way as matter
behaves and curve the space time \cite{Overduin1998}. 
These terms can be interpreted as if, some extra energy coming from the background. 
If the background of a particle change then the terms on the right hand side will change. 
Therefore, any object sitting on that part of the space-time, will fill a force,  which is originated 
from the background and no need to take any fictitious mathematical force.
Therefore, the Mach's principle is explained. 

In this context we can explain the inertial forces that arise in the thought experiment 
given in Sec.\ref{sec2}. In the $K'$ reference frame the distant objects are rotating. 
Therefore, according to Eq.(\ref{eq:Einstein's tensor}), the $G_{\alpha\beta}$ is nonzero. 
This will give rise to the inertial forces required to balance the equation. Therefore, no  
need to postulate unknown form of the inertial forces that only exist in mathematics. 
At this point we should note that the equations resembles the Brans Dick model \cite{Brans1961}  provided  
the $P_{\alpha\beta}$,  $Q_\beta$, $U$ are taken as zero and $k=0$. Though logically 
and mathematically both the theories are completely different.

\section{Machian Gravity in presence of the source terms }

\subsection{Equation of motion and the field equation}

In absence of external force, a particle follows a geodesic path. 
The equation for a geodesic path is given by 

\begin{equation}
\frac{d^{2}x^{A}}{ds^{2}}-\tilde{\Gamma}_{BC}^{A}\frac{dx^{B}}{ds}\frac{dx^{C}}{ds}=0\,.
\end{equation}

\noindent In the weak field limit, considering an almost flat coordinate system, i.e. 
$\tilde{g}_{AB}={\rm diag}(1,-1,-1,-1,-1)+\tilde{\gamma}_{AB} $, and assuming that the 
field is not time varying and the concerned velocities are much less than $c$, 
some straight forward calculations gives $\frac{d^{2}x^{A}}{dt^{2}}=\frac{1}{2} \partial_A \tilde{\gamma}_{00}$ \cite{Einstein}. 
If we relate it with Newtonian gravity we will get

\begin{equation}
\tilde{\gamma}_{00}=2\varphi\,,\label{eq:gamma-phi}
\end{equation}

\noindent where, $\varphi$ is the potential of the gravitational field. 
To get the value of this potential we need to use the Poission's 
equation which may not be same as that of Newtonian mechanics because the terms in the fifth dimension plays
a major role there. 

Also under the above approximations, $\tilde{R}_{00}=\partial_{A}\partial^{A}\tilde{\gamma}_{00}$ \cite{Einstein}.
For a time independent gravitational field, the time derivative is zero. 
In addition, we consider that in small enough region (solar system scale) we can neglect the effect of the the background 
i.e. can ignore the derivatives with respect to $\zeta$ in comparison to the special derivatives.
Therefore, under these approximation we get $\tilde{R}_{00}=\nabla^{2}\tilde{\gamma}_{00}$.
From Eq.(\ref{eq:gamma-phi}) we also know that under such approximations
$\tilde{\gamma}_{00}=2\varphi$. Also, Poisson equation for Newton's
gravitational law is $\nabla^{2}\varphi=4\pi\rho$. Therefore,
we can relate $\tilde{R}_{00}$ to $8\pi\rho$ under the above assumptions.
Similar to GR we can relate $\tilde{G}_{AB}$ with a tensor $\tilde{T}_{AB}$ as 
$\tilde{G}_{AB}=8\pi \tilde{T}_{AB}$. Here $\tilde{T}_{AB}$ is the five dimensional stress energy tensor. 
 In a small region,
where the background fluctuations can be ignored, the left hand side of the equation
will tend to the Einstein's 4D $\tilde{G}_{\mu\nu}$ and the terms in the fifth dimension will 
be negligibly small and will not contribute to the equation. So if we choose 
$\tilde{T}_{AB}$ in such a way that it under the above assumption 
tends to Einstein's stress-energy tensor then we can get back Einstein's equation. 

Though, it is trivial that $T_{;B}^{AB}=0$, it has deeper implications.
It provides the energy-momentum conservation as $T_{;\mu}^{A\mu}+T_{;4}^{A4}=0$,
i.e. the equation has a contribution from the background, which relates the inertial 
properties of a particle with the background objects as explained by Mach's principle.

It is also straight forward to see that the field equation above can be derived from the 
Lagrangian

\begin{equation}
\mathcal{L}=\sqrt{-\tilde{g}}\tilde{R}\,, \label{eq:field equation}
\end{equation}

\noindent which implies that the theory does not violate any conservation principle.

\subsection{Stress energy tensor for fluid}

The properties of a fluid is measured by its density ($\rho$) and
pressure ($p$). Similar to the General theory of
relativity, the stress energy tensor of a
fluid is defined as 
\begin{equation}
\tilde{T}_{AB}=(\rho+p)\tilde{u}_{A}\tilde{u}_{B}+p\tilde{g}_{AB}\,,\label{eq:stress energy tensor}
\end{equation}

\noindent where $\tilde{u}_{A}$ is the 5-velocity of the fluid. 

It can be seen that, the $\tilde{T}_{4\alpha}$ components of the
stress energy tensor are of order $O(\hbar)$ and $\tilde{T}_{44}$
is of the order $O(\hbar^{2})$. Therefore, in the limit $\hbar\rightarrow0$,
these $T_{4A}$ terms can be approximated as $0$. In that case the
stress energy tensor becomes equivalent to the general relativistic
stress energy tensor.

The 5 conservation equations i.e. $\tilde{T}_{;B}^{AB}=0$ along with
the relation between $p$ and $\rho$ and the equation for line element
i.e. $\tilde{g}_{AB}d\tilde{x}^{A}d\tilde{x}^{B}=0$, provides 
a complete solution to the motion of the fluid, provided
$g_{AB}$ is given. This is because there are total $7$ equations
to satisfy $7$ unknowns, which are $p$, $\rho$, $u_{0}$, $\ldots$,$u_{4}$. 

If $g_{AB}$ are also unknown then the field equation
along with some normalization condition $\sqrt{-\tilde{g}}=1$, can
be brought in. This provides $16$ equations for fixing $15$
independent components of $\tilde{g}_{AB}$. Therefore, the equations
may appear over determined. However, it should also be noted that there
are $5$ equations, $\tilde{G}_{;B}^{AB}=0$, which $\tilde{G}^{AB}$ should 
satisfy. Therefore, there are essentially $11$ independent equations in the field equation. 
From the $11$ equations we need to determine the $15$ independent 
component of $\tilde{g}_{AB}$.It shows that we have 
4 degrees of freedom to choose the five dimensional coordinate system.

\section{Static, Spherically symmetric, Vacuum solution for weak gravitation field}

The field equation for the theory is $\tilde{G}_{AB}=8\pi \tilde{T}_{AB}$, which after some rearrangements can be written as

\begin{equation}
\tilde{R}_{AB}=8\pi \left(\tilde{T}_{AB}-\frac{1}{2}\tilde{g}_{AB}\tilde{T}\right)\,.\label{eq:field equation 1}
\end{equation}

\noindent Under the weak field assumption the only relevant component is the $00$
component, which for vacuum solution gives $\tilde{R}_{00}=0$. Also using standard procedure 
$\tilde{R}_{00}$ can be approximated as
$\partial_{C}\partial^{C}\tilde{\gamma}_{00}$. For the static
solution the time derivative also vanish and the equation becomes

\begin{equation}
\partial_{\zeta}^{2}\tilde{\gamma}_{00}+\partial_{x}^{2}\tilde{\gamma}_{00}+\partial_{y}^{2}\tilde{\gamma}_{00}+\partial_{z}^{2}\tilde{\gamma}_{00}=0\,.\label{eq:laplace equation}
\end{equation}

\noindent Under the assumption of spherical symmetry of the special part, the above equation (Eq.(\ref{eq:laplace equation}))
in polar coordinate can be written as

\begin{equation}
\partial_{\zeta}^{2}(r\tilde{\gamma}_{00})+\partial_{r}^{2}(r\tilde{\gamma}_{00})=0\,.\label{eq:waveequation}
\end{equation}

\noindent Using `separation of variables' and considering $(r\tilde{\gamma}_{00})=R(r)\chi(\zeta)$, we get 

\begin{equation}
\frac{1}{R}\frac{\partial^{2}R}{\partial r^{2}}=-\frac{1}{\chi}\frac{\partial^{2}\chi}{\partial\zeta^{2}}=\lambda^{2}\ldots(say)\,,
\end{equation}

\noindent where, $\lambda$ is a constant. This gives

\begin{equation}
R=P_{1}e^{\lambda r}+P_{2}e^{-\lambda r}
\end{equation}

\noindent and 

\begin{equation}
\chi=Q_{1}\cos(\lambda\zeta)+Q_{2}\sin(\lambda\zeta)\,,
\end{equation}

\noindent where, $P_{1}$, $P_{2}$, $Q_{1}$ and $Q_{2}$ are constants. As explained
before, the term $\tilde{\gamma}_{00}$, under weak-field approximation,
is the Newtonian potential and hence it cannot increase exponentially
with distance. Therefore, taking $P_{1}=0$, we can get 
\begin{equation}
(r\tilde{\gamma}_{00})=S+P_{2}e^{-\lambda r}\left(Q_{1}\cos(\lambda\zeta)+Q_{2}\sin(\lambda\zeta)\right)\,.\label{eq:potential-equation}
\end{equation}

\noindent $S$ is constant, as adding a constant will not affect any of
the calculations. If we consider that over a place (a scale of
the order of a galaxy) the background is almost similar then 
the change in $\zeta$ is really small. Therefore, here we may just
take $\lambda\zeta\sim0$. There is a constant factor of  $\hbar$ multiplied
which is also is very small. So, in this limit $\cos(\lambda\zeta)\rightarrow1$
and $\sin(\lambda\zeta)\rightarrow0$.

Replacing these limiting values in Eq.(\ref{eq:potential-equation})
and substituting $P_2Q_1=2KM$ and $S=2(1+K)M$ and replacing $\tilde{\gamma}_{00} = 2\Phi$
we can get the potential as

\begin{equation}
\Phi=\frac{M}{r}\left[1+K\left(1-e^{-\lambda r}\right)\right]\,.\label{eq:potential}
\end{equation}

In Eq.(\ref{eq:potential}), $\lambda$ and $K$ are the background
dependent quantities. Galactic velocity profiles shows that $\lambda$
is of the order of few $kpc^{-1}$. When $r$ is small, $e^{-\lambda r}\sim1$. 
Therefore, $\Phi$
 takes the form of the actual Newtonian potential i.e. $\Phi=\frac{GM}{r}$.
This gives the Newtonian gravitational equation at the solar system
scale. In the asymptotic limit of $r\rightarrow\infty$, the
exponential term goes to $0$. Hence, for large values of $r$, although the form
of the potential will remain same, it will become $(1+K)$
times that of the Newtonian potential.  Recently, similar form of potential is also used 
by other groups to explain the galactic velocity profile correctly \cite{Moffat2009,Moffat2005,Brownstein2005,Brownstein2005a,Moffat2005a}.

\section{Varifying the theory with observations}
\subsection{Galactic rotation curves}

\begin{figure}
\centering
\includegraphics[trim=1.3cm 9.0cm 1.3cm 9.0cm, clip=true, width=0.31\columnwidth]{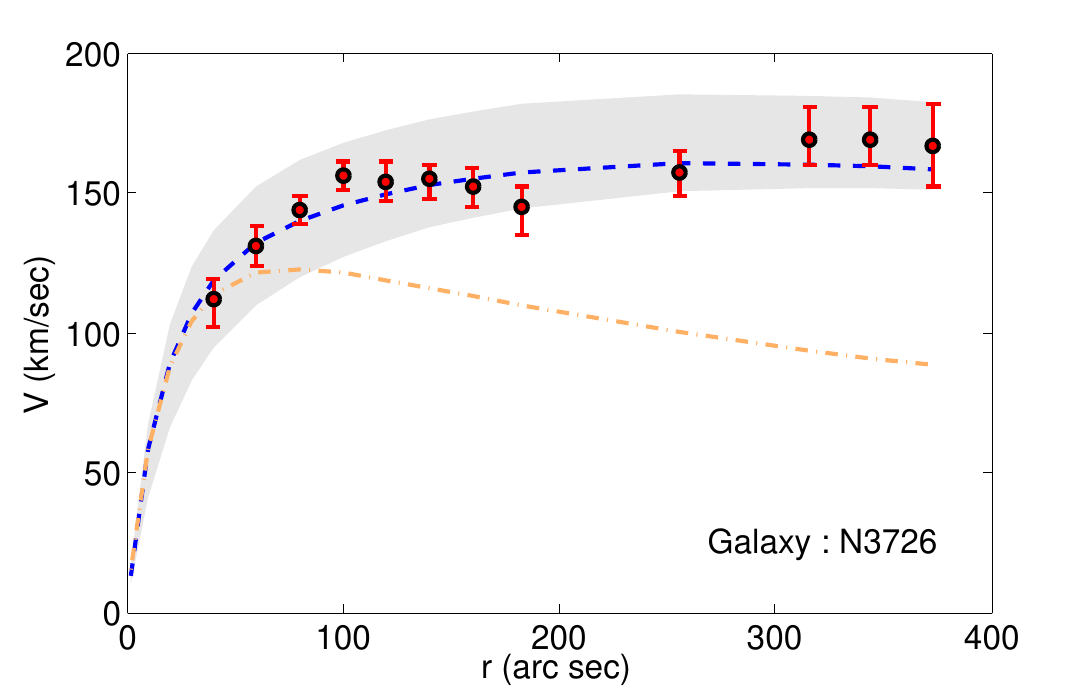}
\includegraphics[trim=1.3cm 9.0cm 1.3cm 9.0cm, clip=true, width=0.31\columnwidth]{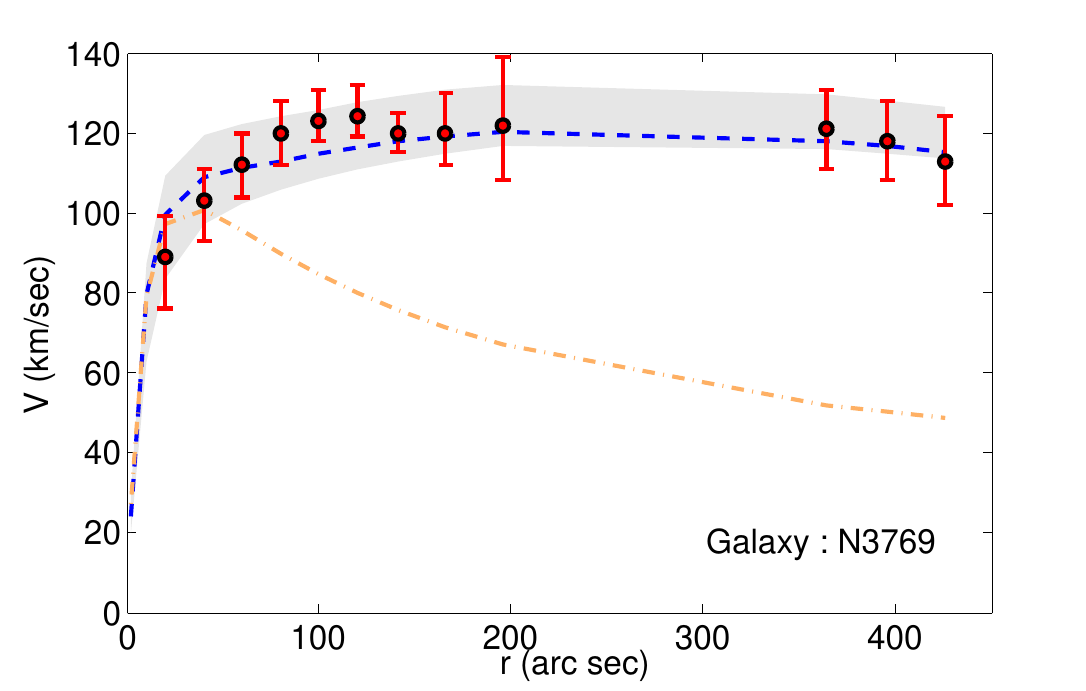}
\includegraphics[trim=1.3cm 9.0cm 1.3cm 9.0cm, clip=true, width=0.31\columnwidth]{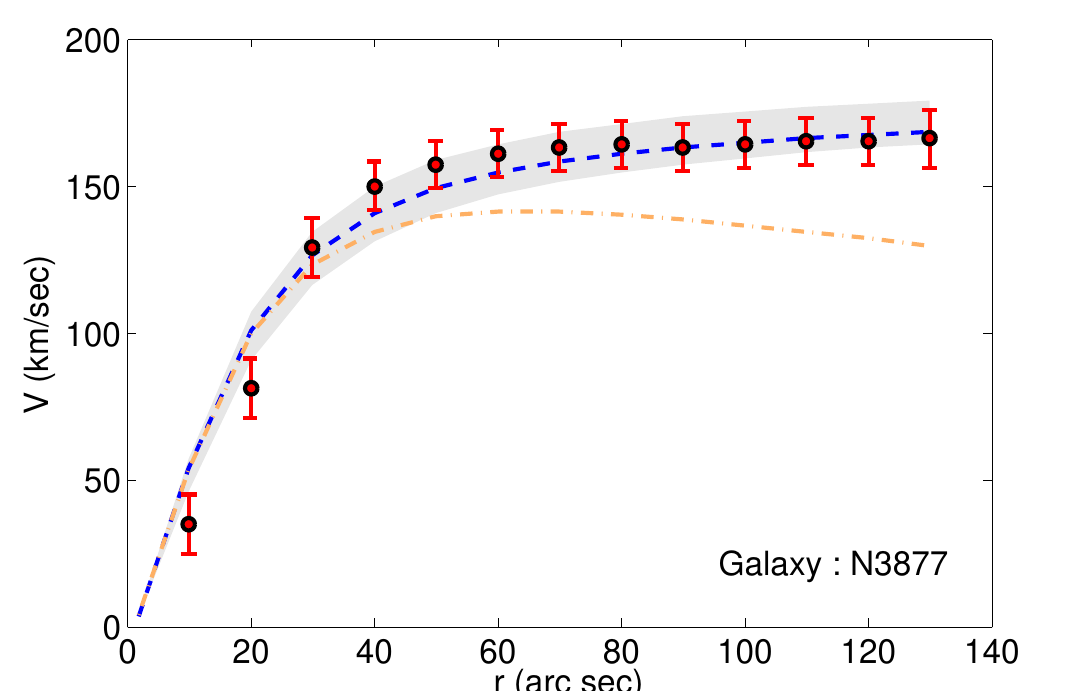}
\includegraphics[trim=1.3cm 9.0cm 1.3cm 9.0cm, clip=true, width=0.31\columnwidth]{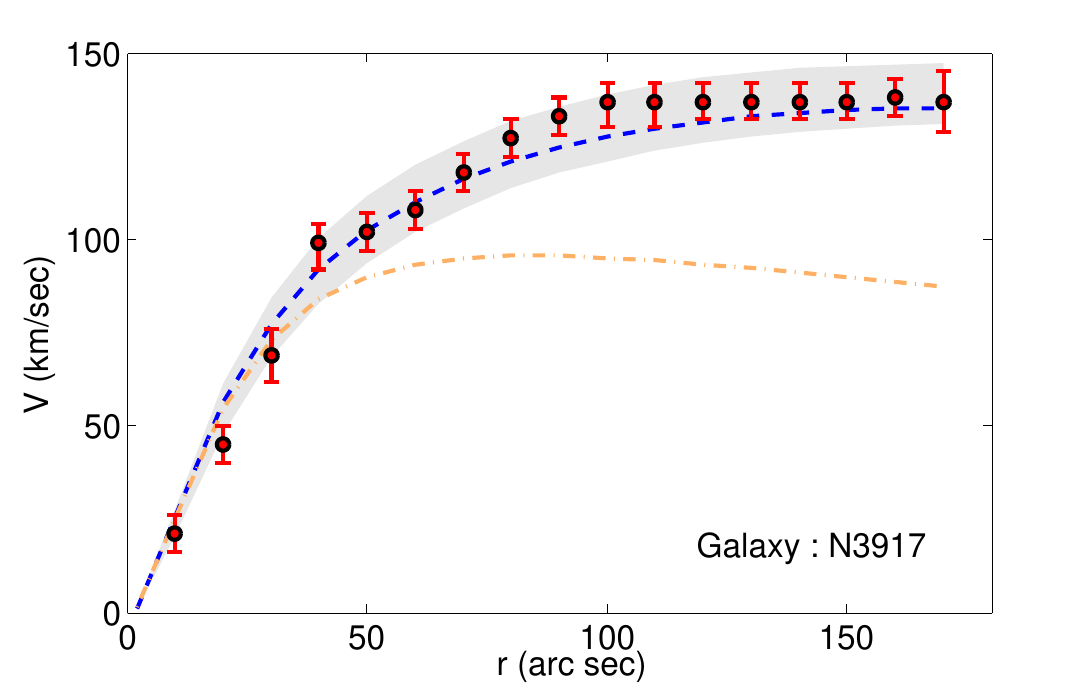}
\includegraphics[trim=1.3cm 9.0cm 1.3cm 9.0cm, clip=true, width=0.31\columnwidth]{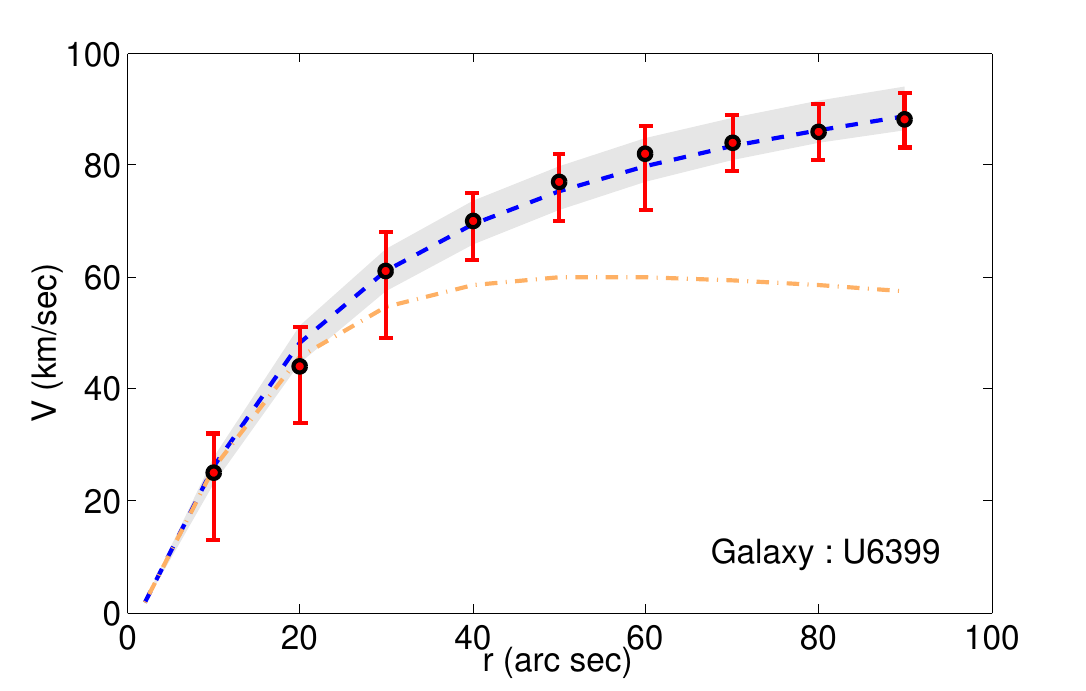}
\includegraphics[trim=1.3cm 9.0cm 1.3cm 9.0cm, clip=true, width=0.31\columnwidth]{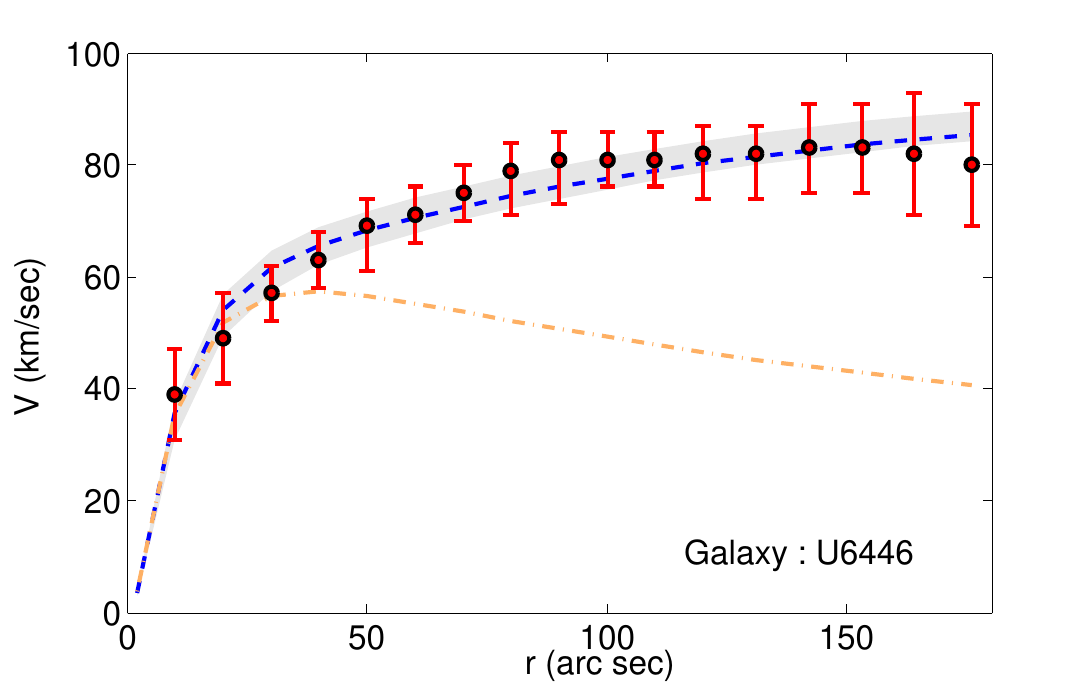}
\includegraphics[trim=1.3cm 9.0cm 1.3cm 9.0cm, clip=true, width=0.31\columnwidth]{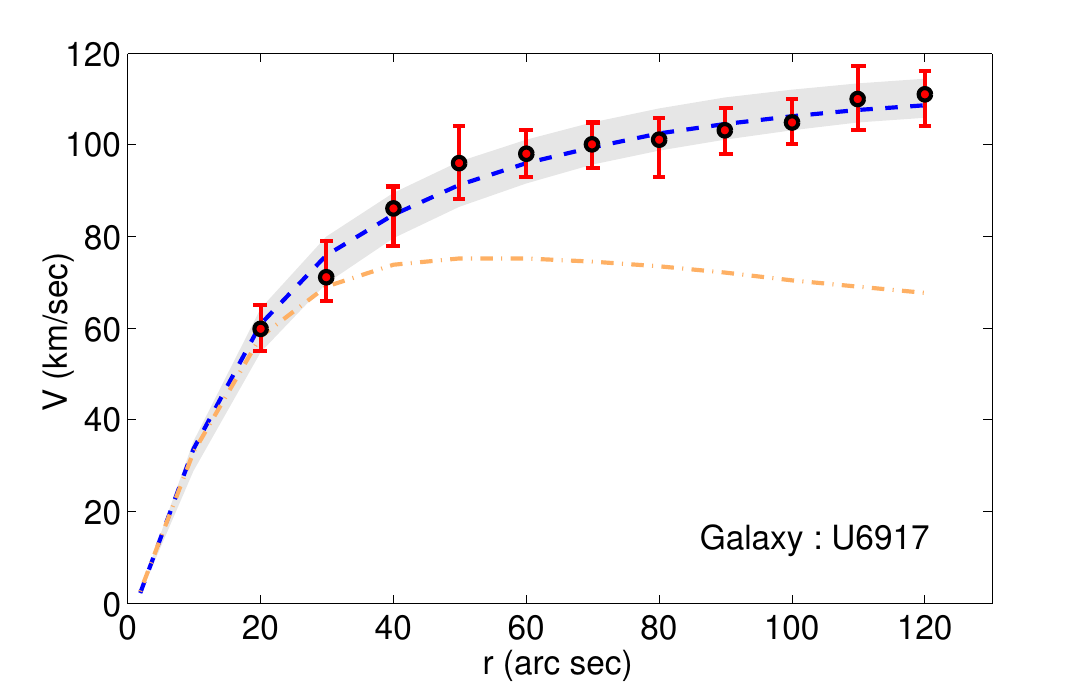}
\includegraphics[trim=1.3cm 9.0cm 1.3cm 9.0cm, clip=true, width=0.31\columnwidth]{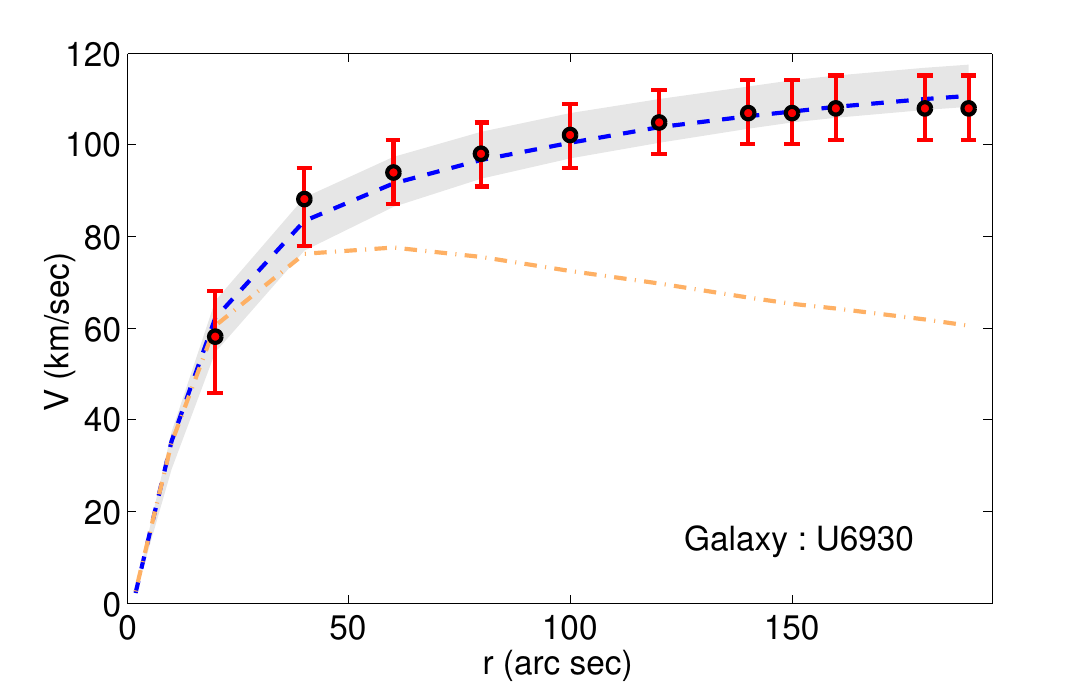}
\includegraphics[trim=1.3cm 9.0cm 1.3cm 9.0cm, clip=true, width=0.31\columnwidth]{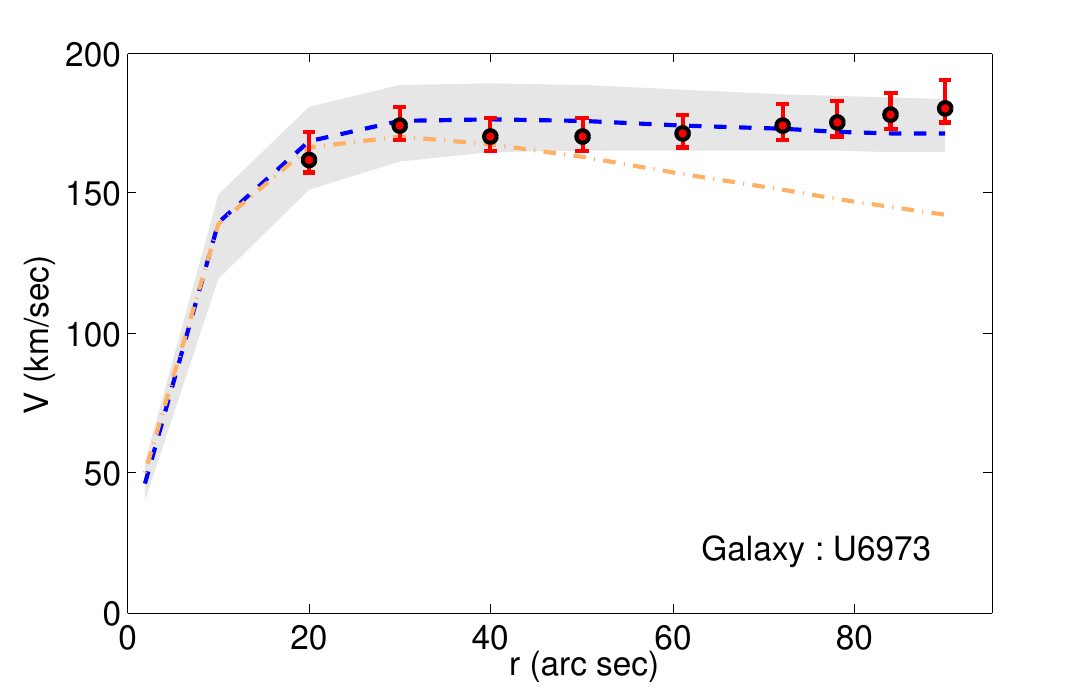}
\includegraphics[trim=1.3cm 9.0cm 1.3cm 9.0cm, clip=true, width=0.31\columnwidth]{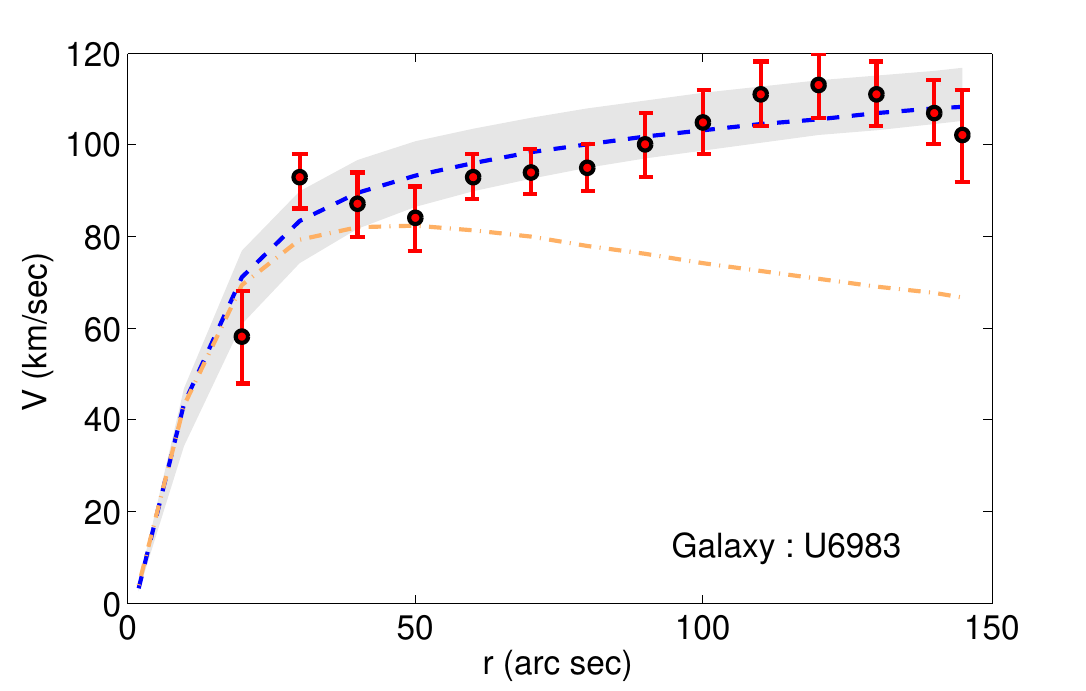}
\includegraphics[trim=1.3cm 9.0cm 1.3cm 9.0cm, clip=true, width=0.31\columnwidth]{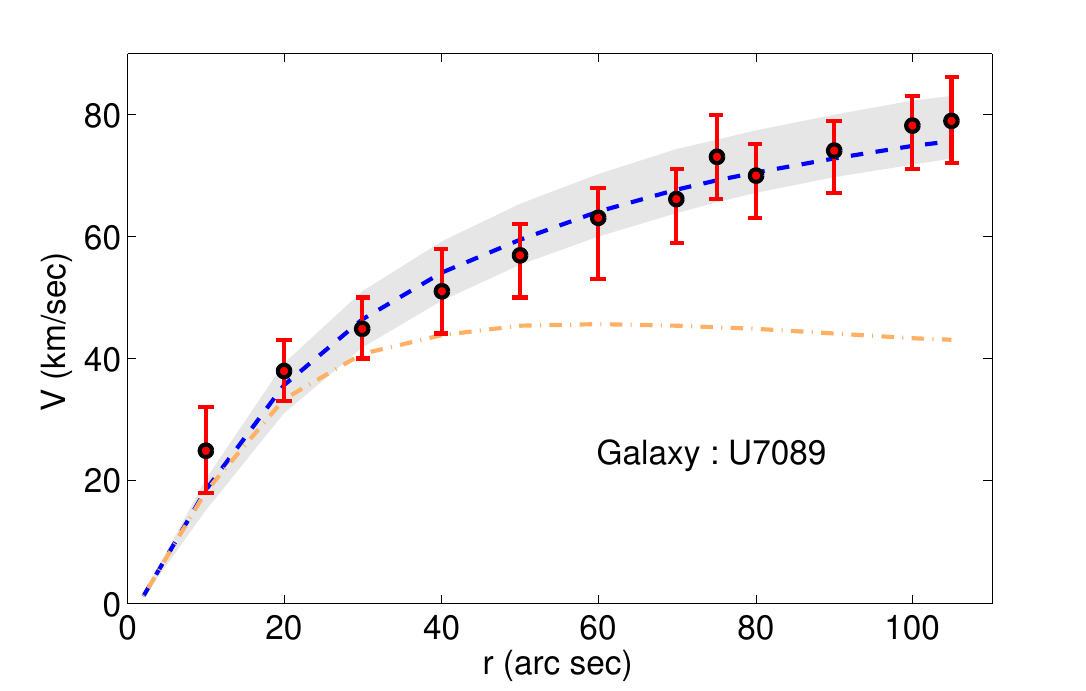}

\caption{\label{fig:GalaxyRotationCurve}The radial velocity profile of some galaxies are plotted with the theoretical values.
The Red plots are showing the Observed data points with the error bars. Blue dotted line is the 
theoretical best fit radial velocity profile from Machian gravity without any dark matter. Gray band is the 
error band on the theoretical value computed from the error bars of the fitted parameters. 
Orange curve is the velocity profile calculated from Newtonian mechanics. In our sample 
NGC3917, U6446, U6917, U6930, U6983, U7089 are the Low Surface Brightness (LSB) galaxies and U6399 is the 
High Surface Brightness (HSB) galaxy. Amongst these U7089 and U6399 are the dwarf galaxies. }

\end{figure}

According to the theory described in this paper the potential due to a static spherically symmetric gravitational 
field is given by Eq.(\ref{eq:potential}). Therefore, the acceleration due to the gravitational field is given by

\begin{equation}
\frac{\partial\Phi}{\partial r}=-\frac{M}{r^{2}}\left[1+K\left(1-e^{-\lambda r}\left(1+\lambda r\right)\right)\right]\,.
\label{eq:Gravfield}
\end{equation}

\noindent Using Eq.(\ref{eq:Gravfield}), the velocity $v$ of an object circulating another object of mass $M$ at 
a distance $r$ can be written as 

\begin{equation}
v=\sqrt{\frac{M}{r}\left[1+K\left(1-e^{-\lambda r}\left(1+\lambda r\right)\right)\right]}\,.
\label{eq:velocity}
\end{equation}

\noindent The equation shows that at large $r$ the velocity of the particles will
be larger than what is expected from the Newton's laws by a factor
of $\sqrt{1+K}$. 

For an observational verification of the theory we fit the galactic rotation curves for a sample 
of 11 galaxies taken from \cite{Verheijen2001}. For theoratical calculation of the 
rotation curve we also need to know mass of the galaxy for different radius which can be calculated by 
measuring the amount of the light and the light mass relation. However, here instead of calculating the 
actual mass we use the parametric fit of the mass \cite{Brownstein2005}, given by $\mathcal{M}(r)=M\left(\frac{r}{r_c+r}\right)^{3\beta}$.
Here $\beta = 1$ for high surface brightness (HSB) galaxies, and $\beta = 2$ for low surface brightness (LSB) galaxies
or dwarf galaxies. $r_c$ is the inner core radius. Where $M$ is the total mass of the galaxy i.e.
 $\rm{lim}_{r\gg r_c} \mathcal{M}(r)=M$. Therefore, the velocity at any radius of the galaxy can be calculated 
by substituting the mass radius relation to Eq.(\ref{eq:Gravfield}). 
The comparison of the radial velocity profile for some galaxies with the 
theoretical value from the theory presented in this paper, are shown in figure (\ref{fig:GalaxyRotationCurve}).
Black dots with red error bars are the observed data points. Dotted blue line is the best fit velovity curve 
calculated using Eq.(\ref{eq:velocity}). We maximize the $\chi^2$ with respect to $M$ and $r_c$ using SCoPE \cite{SCoPE}. 
We use $K = \sqrt{\frac{M_0}{M}}$. 
We also use $\lambda = 0.0718 kpc^{-1}$ and $M_0 = 9.60\times 10^{11}M_{\odot}$
for HSB and LSB galaxies, and  $\lambda = 0.1437 kpc^{-1}$ and  $M_0 = 2.40\times 10^{11}M_{\odot}$ for dwarf galaxies \cite{Brownstein2005}.   
The gray band shows the maximum and the minimum velocities for different combinations of $\{\bar{M}-\Delta M, \bar{M}+\Delta M\}$ and 
$\{\bar{r}_c-\Delta r_c, \bar{r}_c+\Delta r_c\}$, where $\bar{M}$, $\Delta M$, $\bar{r}_c$ and $\Delta r_c$ are the mean and standard deviation 
of $M$ and $r_c$ respectively. Dotted orange curve is the plot from Newtonian mechanics for the best fit values of $M$ and $r_c$.
It can be seen that the theoretical curves fits the observational data points very well. 

\subsection{Galactic cluster mass}

\begin{figure}
\centering
\includegraphics[trim=2.0cm 9.0cm 2.0cm 9.0cm, clip=true, width=0.31\columnwidth]{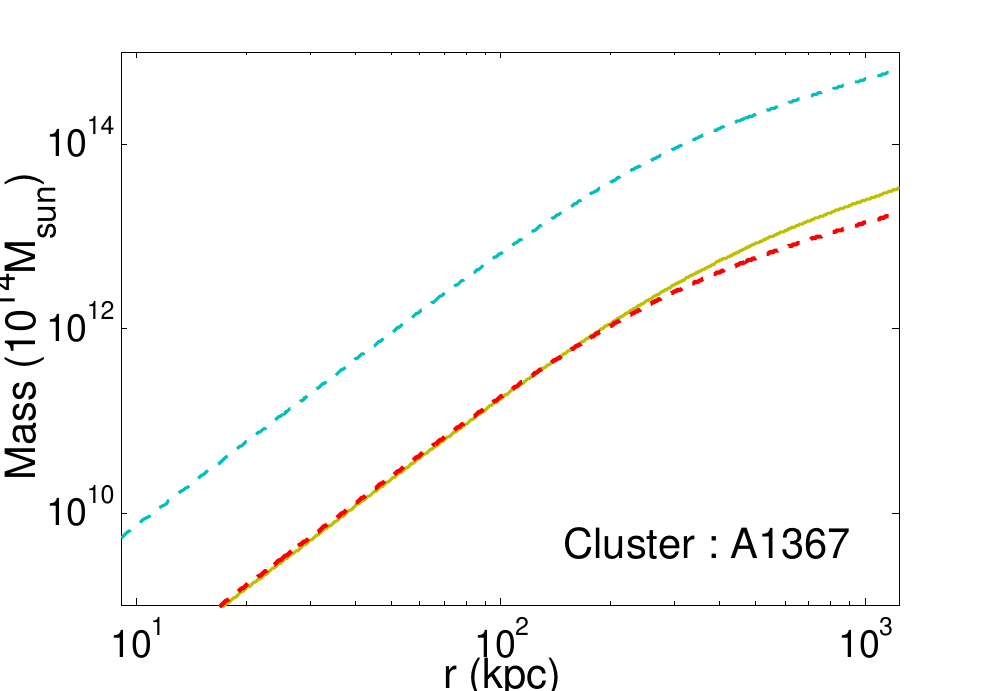}
\includegraphics[trim=2.0cm 9.0cm 2.0cm 9.0cm, clip=true, width=0.31\columnwidth]{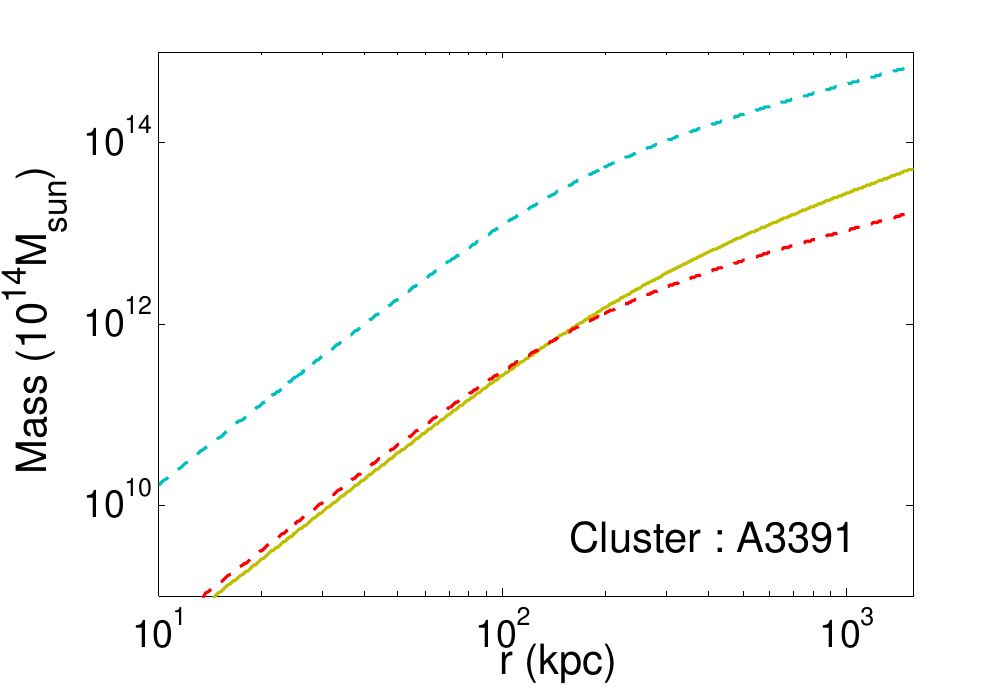}
\includegraphics[trim=2.0cm 9.0cm 2.0cm 9.0cm, clip=true, width=0.31\columnwidth]{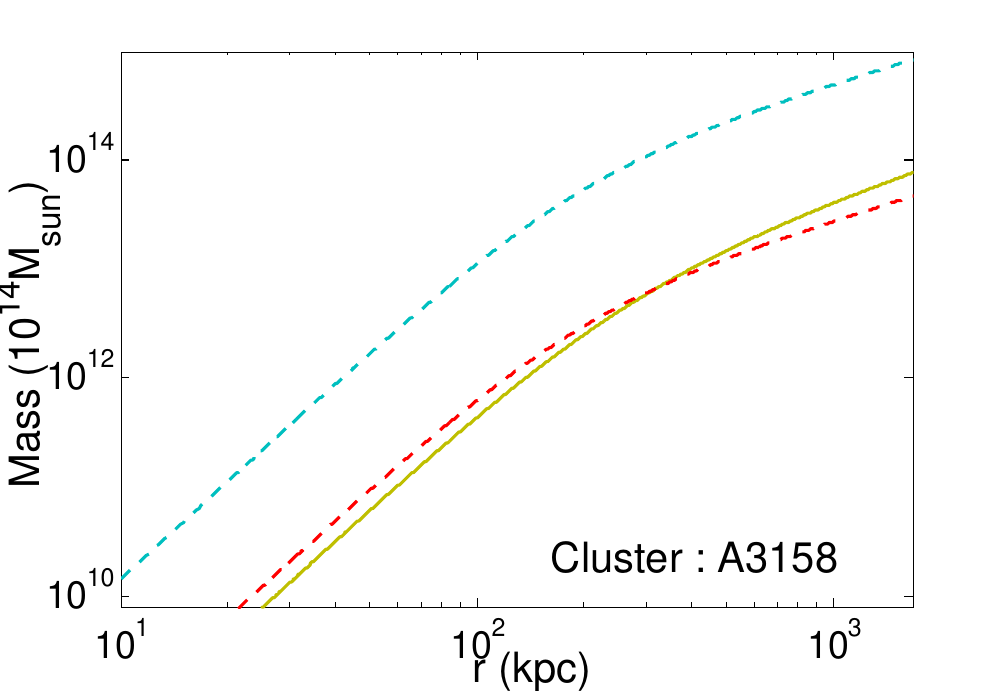}
\includegraphics[trim=2.0cm 9.0cm 2.0cm 9.0cm, clip=true, width=0.31\columnwidth]{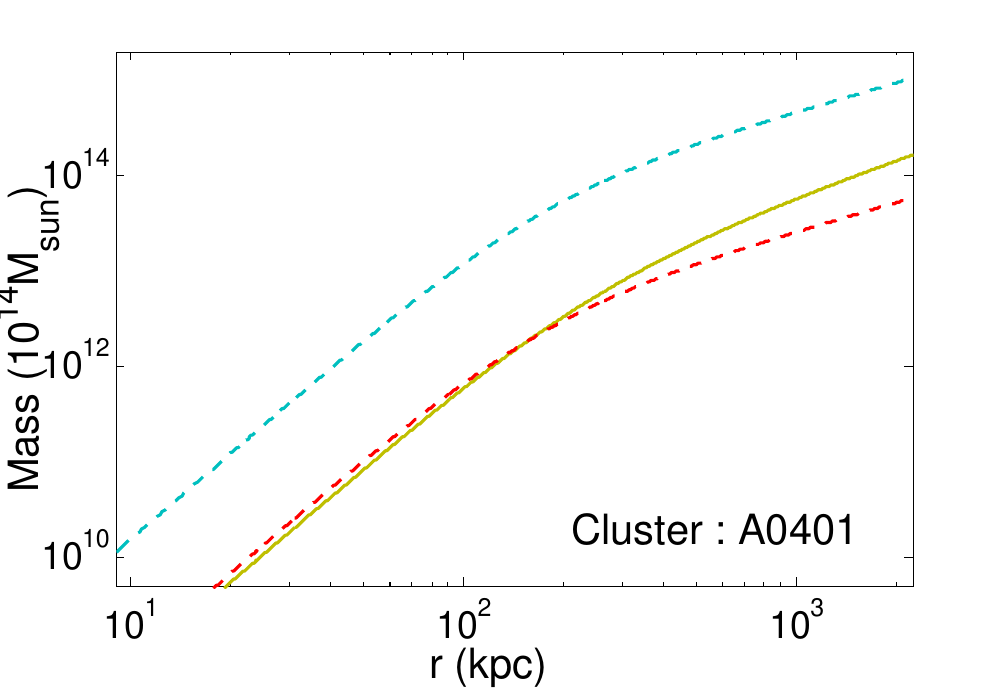}
\includegraphics[trim=2.0cm 9.0cm 2.0cm 9.0cm, clip=true, width=0.31\columnwidth]{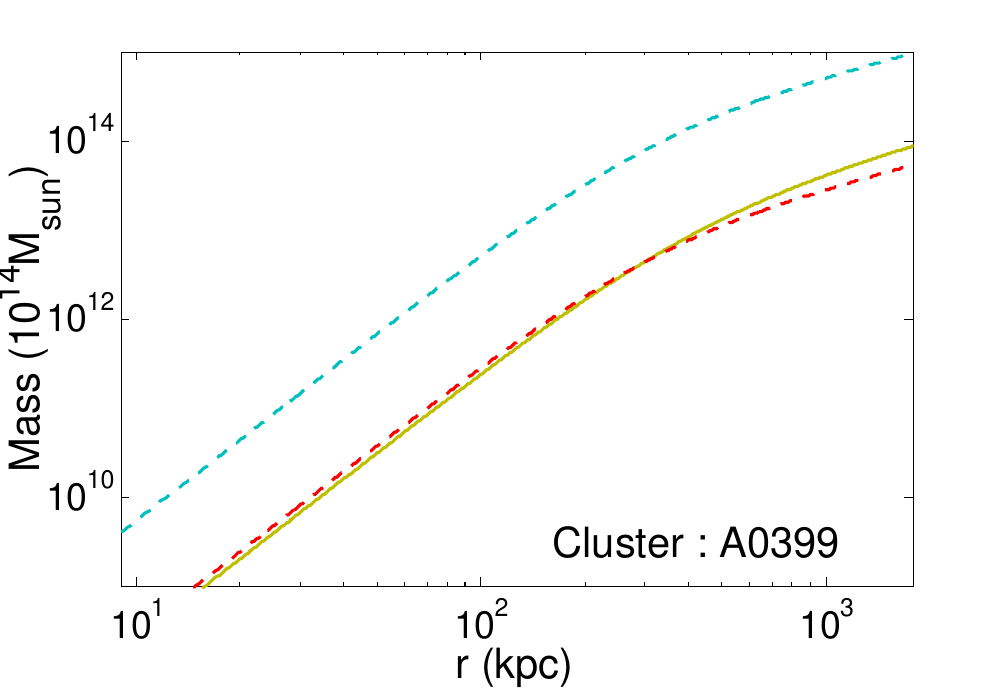}
\includegraphics[trim=2.0cm 9.0cm 2.0cm 9.0cm, clip=true, width=0.31\columnwidth]{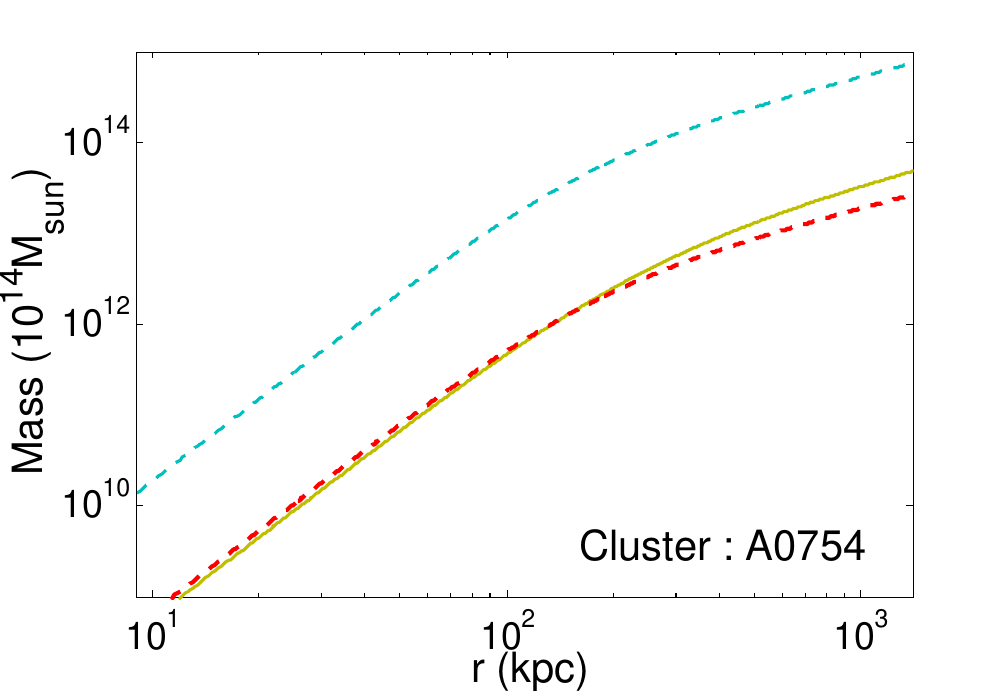}

\caption{\label{fig:GalaxyCusterMass}The radial mass profile of different galaxy clusters are shown here. 
The solid yellow line shows the physical mass of 
the galaxy cluster, calculated using King $\beta$ model. Dotted cyan line shows the dynamic mass of the galaxy cluster calculated given by the 
Newtonian mechanics using Eq.(\ref{eq:M_N}) and the dotted red line shows the dynamic mass of the galaxy cluster given by the theory 
proposed in this paper i.e. using Eq.(\ref{eq:M_M}). The plots show that the dynamic mass from the present theory matches well with the 
physical gas mass of the cluster eliminating requirements of having any exotic dark matter in the clusters.}
\end{figure}

The masses of the $\rm{X_{ray}}$ clusters are another important observational fact that need the exotic dark 
matter candidate to explain its mass content.Here in this section we have shown how can we explain the
galaxy cluster masses using the theory proposed in the chapter without any additional dark matter candidate.

The density of the physical gas in a galaxy cluster can be calculated using the King $\beta$-model
 \cite{King1966,Brownstein2005a,Cavaliere1976}. 

\begin{equation}
\rho(r) = \rho_0 \left[ 1+\left(\frac{r}{r_c}\right)^2\right]^{-3\beta/2}\;,
\label{eq:Kingbeta}
\end{equation}

\noindent where $\beta$ and $r_c$ are parameters which can be calculated from the observed surface 
brightness of the cluster \cite{Brownstein2005a}.
The mass profile of the galaxy can be calculated by integrating the density profile, i.e. 
$M_{King}(r)=4\pi\int_0^r\rho(r')r'^2dr'$. Also, using the velocity dispersion relation due to the temperature
we can find out the dynamical mass of the cluster using Newton's theory. The dynamical mass profile of a 
cluster is given by \cite{Brownstein2005a}

\begin{equation} 
M_N(r) = \frac{3\beta kT}{\mu m_pG_0}\left(\frac{r^3}{r^2+r_c^2}\right)\,,
\label{eq:M_N}
\end{equation}

\noindent where $T$ is the temperature of the cluster, $m_p$ is the mass of proton, $\mu=0.609$ is mean atomic weight
and $k$ is the Boltzmann constant. $G_0$ is the Newton's gravitational constants. However, using the 
Machian gravity theory, proposed in this paper we can write the dynamic mass of the galaxy cluster as 

\begin{equation} 
\mathcal{M}(r) = M_N(r)/\left\{1+K\left[1-\exp(-\lambda r)\left(1+\lambda r\right)\right]\right\}\,.
\label{eq:M_M}
\end{equation}

\noindent As the background of the objects are changing %when we come outside of the cluster from the center of the cluster 
the background of the system will change significantly from the center to the edge and hence $K$ and $\lambda$ should also change
with the radius i.e $r$. However, here for simplicity we use radius independent values $K$ and $\lambda$ . We fix $\lambda = 0.07kpc^{-1}$
for all the clusters and only change $K$.

For testing our theory with the observational results we pick up $6$ galaxy clusters and match the dynamical mass profile 
from our theory with the physical mass given by King's $\beta$-model. We fix the values of $K$ just by eye estimation and 
not by any parameter estimation package. The plots shows that the dynamical mass 
calculated using our theory (shown in orange color)  matches well with the gas mass calculated using the King's $\beta$-model 
(show in yellow ocher). However, using Newtonian 
theory of gravity (cyan dotted lines) we need more dynamical masses than observed, i.e. we need the dark matter. Recently it is seen that  
the MOND \cite{Milgrom1983,Sanders2002} also doesn't provide proper explanation for the galaxy cluster masses. Therefore, it can be said 
that the new theory i.e. Machian Gravity (MG) is better then others alternate theories in all prospects.

\section{Discussion and Conclusion}

A new theory of gravitation, based on Mach's principle is proposed in this paper. It is a metric theory and can be derived from the action principle, 
which guarantees to follow all the conservation principles. The General theory of Relativity or Newtonian gravity are only directly applicable for inertial  observer. In a non inertial or accelerated reference frame we need to add the effect of the acceleration from outside. However, the Machian 
Gravity theory is directly applicable in all the reference frames.  We show that the new theory of gravity can explain the galactic rotation curves 
and the galaxy cluster mass profile very well without any additional dark matter candidate. The lensing of Bullet cluster is 
another phenomenon that can not be explained using GR provided we don't take extra dark matter candidate. 
However, the new theory can explain the lensing of the Bullet Cluster using only physical gas mass of the cluster.  In \cite{Brownstein2007} 
Brownstein and Moffat explained the lensing of the bullet cluster using a gravity equation that is used that is similar to 
Eq.(\ref{eq:Gravfield}). Therefore, the theory can explain different observational results that makes it more credible than other 
standard gravity theories.

\section*{Acknowledgment}
I wish to thank Krishnamohan Parattu, Prof. J. V. Narlikar and Prof. Tarun Souradeep for carefully going through the paper 
and for several interesting discussions.

\bibliographystyle{JHEP}
\bibliography{reference}

\end{document}